\documentclass[twocolumn]{aastex631}

\submitjournal{ApJ Letters}

\usepackage{epstopdf}
\usepackage{graphicx}
\usepackage{threeparttable}
\usepackage{longtable}

\begin{document}

\title{GOALS-JWST: Mid-Infrared Spectroscopy of the Nucleus of NGC 7469}

\correspondingauthor{L. Armus}
\email{lee@ipac.caltech.edu}

\author[0000-0003-3498-2973]{L. Armus}
\affiliation{IPAC, California Institute of Technology, 1200 E. California Blvd., Pasadena, CA 91125, USA}

\author[0000-0001-8490-6632]{T. Lai}
\affiliation{IPAC, California Institute of Technology, 1200 E. California Blvd., Pasadena, CA 91125, USA}

\author[0000-0002-1912-0024]{V. U}
\affiliation{Department of Physics and Astronomy, 4129 Frederick Reines Hall, University of California, Irvine, CA 92697, USA}

\author[0000-0003-3917-6460]{K. L. Larson}
\affiliation{AURA for the European Space Agency (ESA), Space Telescope Science Institute, 3700 San Martin Drive, Baltimore, MD 21218, USA}

\author[0000-0003-0699-6083]{T. Diaz-Santos}
\affiliation{Institute of Astrophysics, Foundation for Research and Technology-Hellas (FORTH), Heraklion, 70013, Greece}
\affiliation{School of Sciences, European University Cyprus, Diogenes street, Engomi, 1516 Nicosia, Cyprus}

\author[0000-0003-2638-1334]{A.S. Evans}
\affiliation{National Radio Astronomy Observatory, 520 Edgemont Rd, Charlottesville, VA, 22903, USA}
\affiliation{Department of Astronomy, University of Virginia, 530 McCormick Road, Charlottesville, VA 22903, USA}

\author[0000-0001-6919-1237]{M. A. Malkan}
\affiliation{Department of Physics \& Astronomy, 430 Portola Plaza, University of California, Los Angeles, CA 90095, USA}

\author[0000-0002-5807-5078]{J. Rich}
\affiliation{The Observatories of the Carnegie Institution for Science, 813 Santa Barbara Street, Pasadena, CA 91101}

\author[0000-0001-7421-2944]{A. M. Medling}
\affiliation{Department of Physics \& Astronomy and Ritter Astrophysical Research Center, University of Toledo, Toledo, OH 43606,USA}
\affiliation{ARC Centre of Excellence for All Sky Astrophysics in 3 Dimensions (ASTRO 3D); Australia}

\author[0000-0002-9402-186X]{D. R.~Law}
\affiliation{Space Telescope Science Institute, 3700 San Martin Drive, Baltimore, MD 21218, USA}

\author[0000-0003-4268-0393]{H. Inami}
\affiliation{Hiroshima Astrophysical Science Center, Hiroshima University, 1-3-1 Kagamiyama, Higashi-Hiroshima, Hiroshima 739-8526, Japan}

\author[0000-0002-2713-0628]{F. Muller-Sanchez}
\affiliation{Department of Physics and Materials Science, The University of Memphis, 3720 Alumni Avenue, Memphis, TN 38152, USA}

\author[0000-0002-2688-1956]{V. Charmandaris}
\affiliation{Department of Physics, University of Crete, Heraklion, 71003, Greece}
\affiliation{Institute of Astrophysics, Foundation for Research and Technology-Hellas (FORTH), Heraklion, 70013, Greece}
\affiliation{School of Sciences, European University Cyprus, Diogenes street, Engomi, 1516 Nicosia, Cyprus}

\author[0000-0001-5434-5942]{P. van der Werf}
\affiliation{Leiden Observatory, Leiden University, PO Box 9513, 2300 RA Leiden, The Netherlands}

\author[0000-0002-2596-8531]{S. Stierwalt}
\affiliation{Physics Department, 1600 Campus Road, Occidental College, Los Angeles, CA 90041, USA}

\author[0000-0002-1000-6081]{S. Linden}
\affiliation{Department of Astronomy, University of Massachusetts at Amherst, Amherst, MA 01003, USA}

\author[0000-0003-3474-1125]{G. C. Privon}
\affiliation{National Radio Astronomy Observatory, 520 Edgemont Rd, Charlottesville, VA, 22903, USA}
\affiliation{Department of Astronomy, University of Florida, P.O. Box 112055, Gainesville, FL 32611, USA}

\author[0000-0003-0057-8892]{L. Barcos-Mu\~noz}
\affiliation{Department of Astronomy, University of Virginia, 530 McCormick Road, Charlottesville, VA 22903, USA}

\author[0000-0003-4073-3236]{C. Hayward}
\affiliation{Center for Computational Astrophysics, Flatiron Institute, 162 Fifth Avenue, New York, NY 10010, USA}

\author[0000-0002-3139-3041]{Y. Song}
\affiliation{Department of Astronomy, University of Virginia, 530 McCormick Road, Charlottesville, VA 22903, USA}
\affiliation{National Radio Astronomy Observatory, 520 Edgemont Rd, Charlottesville, VA, 22903, USA}

\author[0000-0002-7607-8766]{P. Appleton}
\affiliation{IPAC, California Institute of Technology, 1200 E. California Blvd., Pasadena, CA 91125}

\author[0000-0002-5828-7660]{S. Aalto}
\affiliation{Department of Space, Earth and Environment, Chalmers University of Technology, 412 96 Gothenburg, Sweden}

\author{T. Bohn}
\affiliation{Hiroshima Astrophysical Science Center, Hiroshima University, 1-3-1 Kagamiyama, Higashi-Hiroshima, Hiroshima 739-8526, Japan}

\author[0000-0002-5666-7782]{T. B\"oker}
\affiliation{European Space Agency, Space Telescope Science Institute, Baltimore, MD 21218, USA}

\author[0000-0002-1207-9137]{M. J. I. Brown}
\affiliation{School of Physics and Astronomy, Monash University, Clayton, VIC 3800, Australia}

\author[0000-0002-1392-0768]{L. Finnerty}
\affiliation{Department of Physics \& Astronomy, 430 Portola Plaza, University of California, Los Angeles, CA 90095, USA}

\author[0000-0001-6028-8059]{J. Howell}
\affiliation{IPAC, California Institute of Technology, 1200 E. California Blvd., Pasadena, CA 91125}

\author[0000-0002-4923-3281]{K. Iwasawa}
\affiliation{Institut de Ci\`encies del Cosmos (ICCUB), Universitat de Barcelona (IEEC-UB), Mart\'i i Franqu\`es, 1, 08028 Barcelona, Spain}
\affiliation{ICREA, Pg. Llu\'is Companys 23, 08010 Barcelona, Spain}

\author[0000-0003-2743-8240]{F. Kemper}
\affiliation{Institut de Ciencies de l'Espai (ICE, CSIC), Can Magrans, s/n, 08193 Bellaterra, Barcelona, Spain}
\affiliation{ICREA, Pg. Lluís Companys 23, Barcelona, Spain}
\affiliation{Institut d'Estudis Espacials de Catalunya (IEEC), E-08034 Barcelona, Spain}

\author[0000-0001-7712-8465]{J. Marshall}
\affiliation{Glendale Community College, 1500 N. Verdugo Rd., Glendale, CA 91208}

\author[0000-0002-8204-8619]{J. M. Mazzarella}
\affiliation{IPAC, California Institute of Technology, 1200 E. California Blvd., Pasadena, CA 91125}

\author[0000-0002-6149-8178]{J. McKinney} 
\affiliation{Department of Astronomy, University of Massachusetts, Amherst, MA 01003, USA.}

\author[0000-0001-7089-7325]{E.J.\,Murphy}
\affiliation{National Radio Astronomy Observatory, 520 Edgemont Road, Charlottesville, VA 22903, USA}

\author[0000-0002-1233-9998]{D. Sanders}
\affiliation{Institute for Astronomy, University of Hawaii, 2680 Woodlawn Drive, Honolulu, HI 96822}

\author[0000-0001-7291-0087]{J. Surace}
\affiliation{IPAC, California Institute of Technology, 1200 E. California Blvd., Pasadena, CA 91125}


\begin{abstract}
We present mid-infrared spectroscopic observations of the nucleus of the nearby Seyfert galaxy NGC 7469 taken with the MIRI instrument on the \emph{James Webb Space Telescope (JWST)} as part of Directors Discretionary Time Early Release Science (ERS) program 1328. The high resolution nuclear spectrum contains 19 emission lines covering a wide range of ionization. The high ionization lines show broad, blueshifted emission reaching velocities up to 1700 km s$^{-1}$ and FWHM ranging from $\sim500 - 1100$ km s$^{-1}$. The width of the broad emission and the broad to narrow line flux ratios correlate with ionization potential. The results suggest a decelerating, stratified, AGN driven outflow emerging from the nucleus. The estimated mass outflow rate is one to two orders of magnitude larger than the current black hole accretion rate needed to power the AGN. Eight pure rotational H$_{2}$ emission lines are detected with intrinsic widths ranging from FWHM $\sim 125-330$ km s$^{-1}$. We estimate a total mass of warm H$_{2}$ gas of $\sim1.2\times10^{7}$M$_{\odot}$ in the central 100 pc. The PAH features are extremely weak in the nuclear spectrum, but a $6.2\mu$m PAH feature with an equivalent width $\sim0.07\mu$m and a flux of $2.7\times10^{-17}$ W m$^{-2}$ is detected. The spectrum is steeply rising in the mid-infrared, with a silicate strength $\sim0.02$, significantly smaller than seen in most PG QSOs, but comparable to other Seyfert 1's. These early MIRI mid-infrared IFU data highlight the power of \emph{JWST} to probe the multi-phase interstellar media surrounding actively accreting supermassive black holes.

\end{abstract}


\section{Introduction}

Luminous Infrared Galaxies (LIRGs), with thermal IR[8-1000~$\mu$m] dust emission in excess of 10$^{11}$~L$_\odot$, are ideal laboratories for studying star formation and black hole growth in the local Universe. 
The bolometric luminosity of most LIRGs is dominated by massive bursts of star formation, but also show a wide range of contributions from Active Galactic Nuclei (AGN) which become increasingly important at the highest luminosities \citep{petric011, Stierwalt13, Stierwalt14}. Multi-wavelength observations have shown that local LIRGs are a mixture of single disk galaxies, interacting systems, and advanced mergers \citep{Stierwalt13, kim2013, larson16}. 

Our understanding of the energetics and evolutionary states of LIRGs at low and high-redshifts was greatly expanded through studies with the Infrared Spectrograph \citep[IRS - see][]{houck04} on the \emph{Spitzer Space Telescope} \citep[see][for a summary]{armus20}.  While \emph{Spitzer} was extremely efficient and capable of deep as well as large area photometric and spectroscopic surveys, it was ultimately limited in sensitivity and spatial resolution by its modest 0.85m primary mirror. With the successful launch and commissioning of the \emph{James Webb Space Telescope (JWST)}, and its 1-2 orders of magnitude gain in both sensitivity, and spatial and spectral resolution over \emph{Spitzer}, we are now poised to greatly extend our understanding of dusty star-forming regions, nascent AGN, galactic outflows, and all variety of active galaxies over an extremely large stretch of cosmic time. To highlight the power of \emph{JWST} to explore the inner regions of nearby LIRGs, we have undertaken a Director's Discretionary Time, Early Release Science (ERS) program (1328 - PI's Armus, L., Evans, A.S.) to obtain near and mid-infrared images and spectra of four LIRGs drawn from the Great Observatories All-Sky LIRG Survey \citep[GOALS -][]{armus09}.  GOALS uses multi-wavelength observations from the ground and space to understand the physical conditions in over 200 low-redshift ($ z < 0.09$) LIRGs selected from the Revised IRAS Bright Galaxy Sample \citep{sanders03}.  This paper highlights some of the first results from ERS program 1328, ``A \emph{JWST} Study of the Starburst-AGN Connection in Merging Luminous Infrared Galaxies".


NGC 7469 (IRAS 23007+0836; UGC 12332) contains a Seyfert 1.5 nucleus surrounded by a bright, circumnuclear ring of star formation with a radius of 1.8” (580 pc). The total infrared luminosity of NGC 7469 is $4.5\times10^{11}$ L$_{\odot}$. It is a nearly face-on spiral, about 26 kpc away in projection from an inclined companion galaxy (IC 4283) with a highly disturbed morphology. NGC 7469 has a supermassive black hole mass of $\sim10^7$M$_{\odot}$, measured through reverberation mapping \citep{peterson04}. The starburst ring is visible at multiple wavelengths, with a large number of star forming regions having a bi-modal age distribution \citep{Diaz-Santos07}. There is evidence for a highly ionized, outflowing wind on small and large scales \citep{U19, Muller-Sanchez11, Robleto-Orus21}. The mid-infrared \emph{Spitzer} spectrum of NGC 7469 shows a mix of strong Poly-cyclic Aromatic Hydrocarbon (PAH) emission features, high-ionization lines, and warm molecular gas, indicative of a complex, composite source powered by an active nucleus and a starburst \citep{Inami13, Stierwalt14}. High resolution observations of CO($1-0$), CO($2-1$) and [CI] in the inner $\sim$ 2 kpc region with ALMA show largely rotational kinematics and a large mass of molecular gas \citep{Izumi20,Nguyen21}. With a central AGN, a circumnuclear starburst ring, a high velocity outflow, and a dusty, gas rich ISM, NGC 7469 is an ideal laboratory for studying the co-evolution of galaxies and supermassive black holes at high resolution with \emph{JWST}.
 
Here, we present \emph{JWST} integral field spectroscopic observations of the nucleus of NGC 7469 with the Mid-Infrared Instrument \cite[MIRI;][]{Rieke15,Labiano21}. The high spectral-spatial resolution enables us to map the physical properties and kinematics of the atomic and warm molecular gas as well as the dust on sub-kpc scales in the mid-infrared for the first time.  Specifically, the Integral Field Unit (IFU) data allow us to separately extract and analyze high signal-to-noise, mid-infrared spectra of the starburst ring, the inner ISM, and the central AGN. The ring spectra are presented in \cite{lai22}.  Spectra of the circumnuclear ISM are presented in \cite{u22}.  Here, we focus on the MIRI spectra of the central active nucleus. Throughout this paper, we adopt $H_0 = 70$\,km\,s$^{-1}$\,Mpc$^{-1}$, $\Omega_{\rm m}$ = 0.30, and $\Omega_{\rm vac}$ = 0.70. At the distance of NGC 7469, 70.6 Mpc ($z = 0.01627$), 1\arcsec  subtends a projected linear scale of 330 pc. 

\section{Observations and Data Reduction}

Observations of NGC 7469 were taken with MIRI in Medium Resolution Spectroscopy (MRS) mode on 2022 July 3-4 UT. The observations covered the full $4.9-28.8\mu$m~range using the short (A), medium (B), and long (C) sub-bands in all four channels. The FASTR1 readout pattern was used to optimize the dynamic range in the observations. The total exposure time per sub-band was 444 seconds, using 40 groups per integration and a 4-pt dither pattern. Because NGC 7469 is extended, dedicated background observations with the same observational parameters in all three grating settings were obtained. MIRI imaging observations were taken using three filters, F560W, F770W, and F1500W, using a cycling three point dither pattern. Data were collected with both the full imaging field of view (74\arcsec $\times 113$\arcsec) and with the SUB128 subarray mode (14.1\arcsec $\times 14.1$\arcsec) to recover a non-saturated image of the bright nucleus. The total exposure times are 309s, 101s, and 101s and 46s, 48s, and 48s for the F560W, F770W, F1500W filters and full and sub-array modes respectively. The MIRI and NIRCam imaging of the starburst ring in NGC 7469 will be presented in \cite{bohn22}.

Data for the science and background observations were processed with the \emph{JWST} Science Calibration Pipeline \cite{jwstpipe} version 1.6$+$ in batch mode. The Detector1 pipeline applies detector-level corrections and ramp fitting to the individual exposures. The output rate images were subsequently processed outside the \emph{JWST} pipeline to flag newly-acquired bad pixels and additional cosmic ray artifacts, and to remove vertical stripes and zero point residuals remaining after the pipeline dark subtraction.  These additional corrections broadly follow the steps taken for \emph{JWST} ERO observations as described by \citet{pontop22}.  The resulting rate files are then processed with the \emph{JWST} Spec2 pipeline for distortion and wavelength calibration, flux calibration, and other 2d detector level steps. Stage 3 processing (Spec3) performs background subtraction before combining data from multiple exposures into the final data cubes. Background light is subtracted from the 2-D science images using a master background frame generated from our associated background observations. The master background is a 1-D median sigma-clipped spectrum calculated over the FOV of the background observations and projected to the entire 2-D detector array. Residual fringe corrections using prototype pipeline code have been applied to both the Stage 2 products and to the 1-D spectra resulting from Stage 3 processing after spectral extraction (see below).

\begin{figure}
	\centering
	\includegraphics[width=0.47\textwidth]{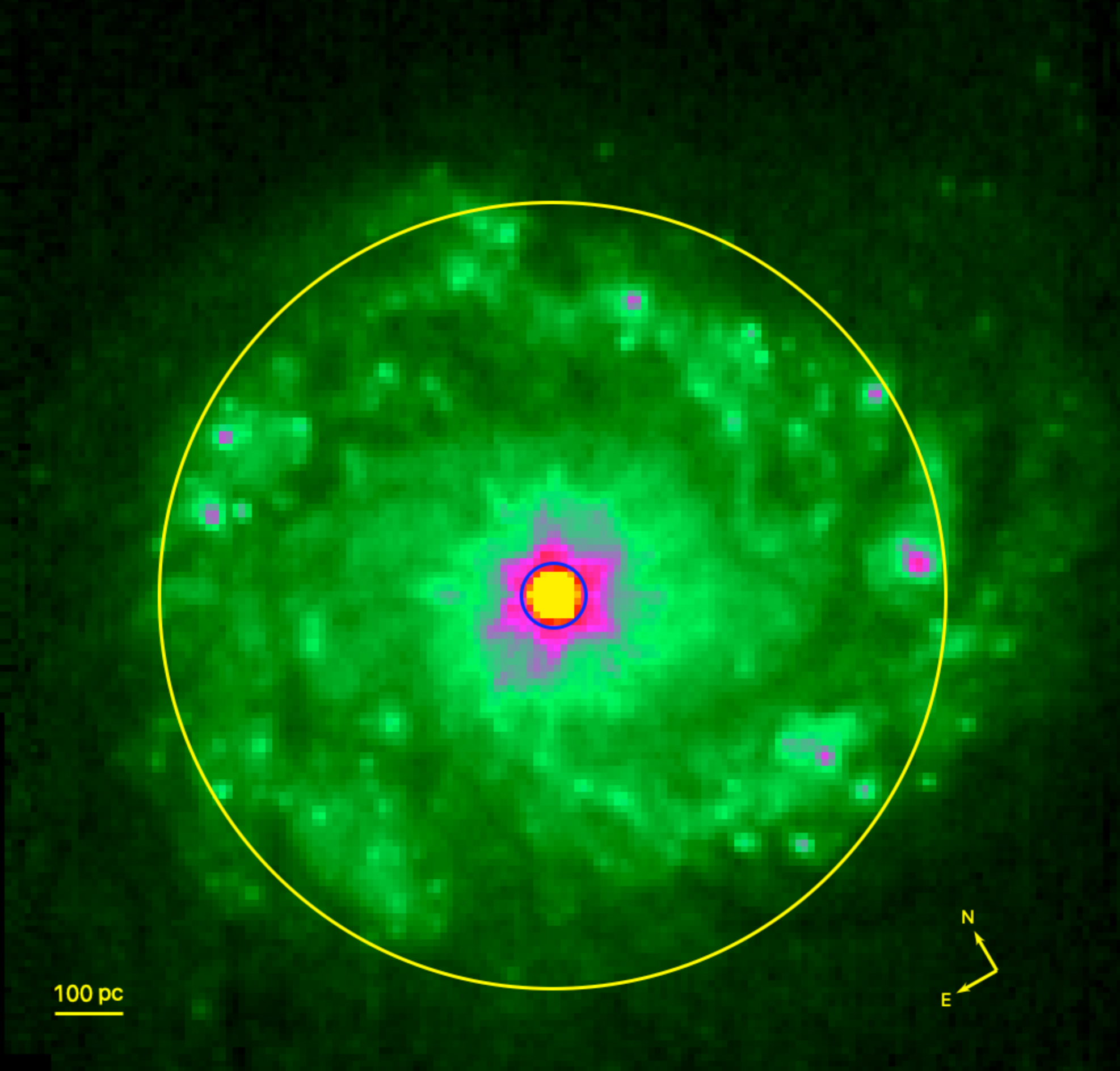}
	\caption{False color \emph{JWST} NIRCam F150W sub-array image of the center of NGC 7469.  The image, which is 5.1\arcsec (1.65 kpc) across, shows the bright, central AGN and the surrounding starburst ring, which is resolved into dozens of star forming knots, interspersed with regions of diffuse emission and dark dust lanes. In the image, two circles are shown that represent the extraction regions for the MIRI total (large yellow circle - 3.6\arcsec diameter) and nuclear (small blue circle - 0.3\arcsec diameter) MRS spectra discussed in the text. The total spectrum includes the AGN and the starburst ring. The nuclear spectrum is dominated by the AGN, and is the focus of this paper.  A scale bar depicting 100 pc in projection is shown at the lower left.}
	\label{FIG:apertures}
\end{figure}

\begin{deluxetable*}{lllll}
\tablecolumns{5} 
\tablecaption{Measured Spectral Features}
 \tablehead{
 \colhead{Feature ID}& 
 \colhead {$\lambda_{rest}$} &
 \colhead {IP} &
 \colhead{Flux} & 
 \colhead{FWHM} \\
 \colhead{}& 
 \colhead{$(\micron$)}&
 \colhead{(eV)} &
 \colhead {$(10^{-18} W$  $m^{-2}$)}&
 \colhead{($km$ $s^{-1}$)}
 }

\startdata
H$_{2}$ S(8)&5.053& &0.64 (0.17)&268 (47)\\
$[FeII]$&5.062&7.9&0.38 (0.07)&$< 100$\\
HeII&5.228&24.6&1.29 (0.19)&298 (33)\\
$[FeII]$&5.340&7.9&1.31 (0.07)&234 (11)\\
$[FeVIII]$&5.447&124&1.98 (0.28)&372 (26)\\
 & & &4.09 (0.36)&1095 (63) \\
$[MgVII]$&5.503&186.5&1.03 (0.32)&189 (32)\\
& & &4.05 (0.42)&621 (36)\\
H$_{2}$ S(7)&5.511& &1.45 (0.15)&124 (18)\\
$[MgV]$&5.610&109.2&3.51 (0.33)&231 (11)\\
& & &9.66 (0.41)&726 (22)\\
H$_{2}$ S(6)&6.11& &0.53 (0.05)&175 (12)\\
6.2 PAH&6.22& &27.1 (0.2)&$--$\\
H$_{2}$ S(5)&6.910& &3.05 (0.11)&188 (5)\\
$[ArII]$&6.985&15.8&13.48 (0.61)&261 (10)\\
$[NaIII]$&7.318&47.3&1.76 (0.17) &427 (29)\\
$[NeVI]$&7.652&126.2&26.7 (1.99)&309 (13)\\
& & &18.0 (2.25)&962 (64)\\
H$_{2}$ S(4)&8.025& &1.59 (0.23)&214 (23)\\
$[ArIII]$&8.991&27.6&7.00 (2.78)&291 (38)\\
& & &5.31 (1.47)&563 (156)\\
$[FeVII]$&9.527&99.1&2.07 (0.69)&237 (34)\\
& & &2.35 (0.74)&631 (197)\\
H$_{2}$ S(3)&9.665& &2.94 (0.63)&265 (43)\\
$[SIV]$&10.511&34.8&22.86 (0.99)&299 (9)\\
& & &10.9 (1.40)&667 (65)\\
H$_{2}$ S(2)&12.279& &2.71 (0.51)&202 (29)\\
$[NeII]$&12.814&21.6&49.62 (3.62)&313 (14)\\
$[NeV]$&14.322&97.1&41.77 (3.64)&259 (13)\\
& & &32.68 (4.51)&814 (62)\\
$[NeIII]$&15.555&41.0&91.61 (4.65)&287 (8)\\
& & &45.85 (4.79)&920 (38) \\
H$_{2}$ S(1)&17.035& &4.16 (1.95)&331 (140)\\
$[FeII]$&17.936&7.9&2.46 (0.97)&319 (90)\\
$[SIII]$&18.713&23.3&46.77 (1.37)&207 (5)\\
$[NeV]$&24.318&97.1&42.48 (4.07)&224 (12)\\
& & &34.82 (6.24)&861 (85)\\
$[OIV]$&25.890&54.9&151.73 (26.31)&261 (20)\\
& & &96.4 (34.61)&706 (165) \\
\enddata

\tablecomments{Basic properties of the emission lines measured in the \emph{JWST/MIRI} MRS nuclear spectrum of NGC 7469. Column 1: Feature ID; Column 2: rest wavelength ($\mu$m); Column 3: line flux ($10^{-18}$ W m$^{-2}$); Column 4: intrinsic FWHM (km s$^{-1}$) after subtraction of the instrumental broadening. Lines that have been fit with an additional broad component have a second entry row. One-sigma uncertainties in the fitted quantities are given in parentheses. The $6.2\mu$m emission feature is a series of resolved components, not a single emission line, so does not have an entry in Column 4.}
\label{TAB:linefits}
\end{deluxetable*}

\begin{figure*}
	\centering
	\includegraphics[width=0.85\textwidth]{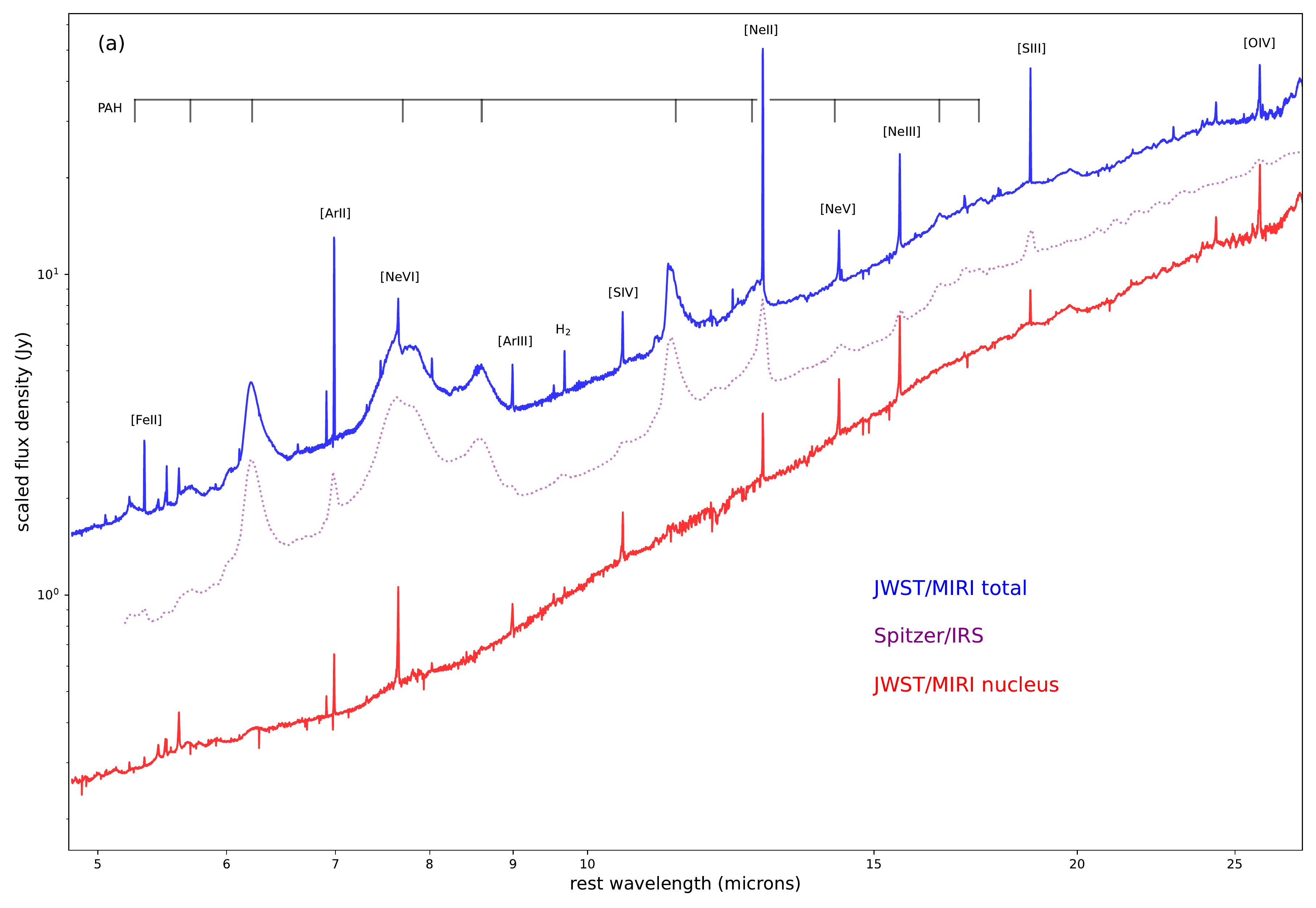}
    \includegraphics[width=0.85\textwidth]{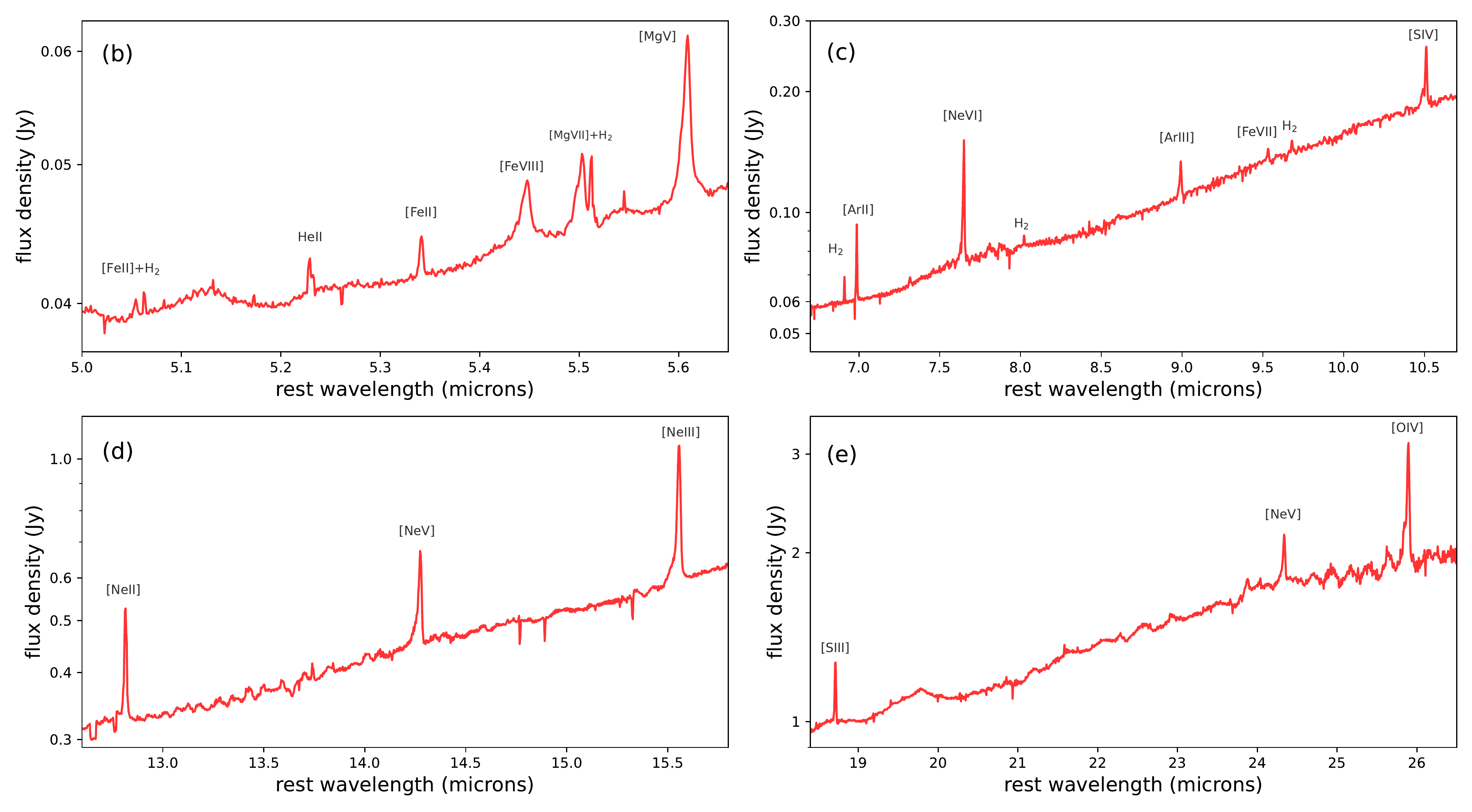}
	\caption{\emph{JWST/MIRI} mid-infrared spectra of NGC 7469. The top panel (a) is a comparison of the total MIRI/MRS spectrum (blue solid) and the nuclear extraction (red solid) to the \emph{Spitzer/IRS} low-res spectrum (purple dotted) from \cite{Stierwalt13}. All spectra are arbitrarily scaled and presented in the rest frame.  Prominent emission lines and PAH features are labeled at the top. Atomic lines covering a wide range of ionization potentials and a number of H$_{2}$ lines from warm molecular gas are present in the NGC 7469 nuclear spectrum, but the PAH features that dominate the total \emph{JWST} and \emph{Spitzer} spectra are weak or absent. The bottom four panels (b-d) highlight four spectral regions of the nuclear spectrum with individual bright emission features labeled.  Asymmetric blue wings and broad profiles are evident on many of the high ionization emission lines, indicative of fast moving gas associated with a nuclear outflow.} 
	\label{FIG:FullSpectra}
\end{figure*}

\section{Results}

We have extracted two spectra from the NGC 7469 MRS data cubes. The first, hereafter the ``total" spectrum, is defined as the largest cylindrical aperture that covers the full wavelength extent of MIRI. This aperture has a radius of 1.8\arcsec, centered on the nucleus. The second extraction, meant to isolate the central AGN, is also centered on the nucleus, but with a one FWHM diameter (0.3\arcsec at $5\mu$m) expanding cone. Because this nuclear aperture is fairly small and dominated by the central AGN by design, a wavelength-dependent aperture correction, constructed from standard star observations has been applied to the data before measuring emission line fluxes. Both apertures are drawn on the NIRCAM F150W sub-array image in Fig.~\ref{FIG:apertures}.

The two NGC 7469 MIRI spectra are shown in Fig.~\ref{FIG:FullSpectra}, along with the low-resolution $Spitzer/IRS$ spectrum from \cite{Stierwalt13} for comparison. The \emph{Spitzer/IRS} spectrum is derived from a single pointing on the nucleus with the short-low (3.6\arcsec wide) and long-low (10.5\arcsec wide) slits. As in the \emph{Spitzer} spectrum, the total NGC 7469 MIRI spectrum is dominated by PAH emission features (complexes) at 6.2, 7.7, 8.6, 11.3 and 12.6$\mu$m. There are also weaker PAH features at 5.3, 14.2, 16.4 and 17.4$\mu$m.  In addition, strong, narrow atomic fine structure emission lines from a variety of species such as Fe, Ne, Ar, Si, S, O, Mg, and Na, are visible.  Emission lines from a number of rotational transitions of warm H$_{2}$ are also seen throughout the spectrum. The brightest of these atomic and molecular gas emission lines were visible in the \emph{Spitzer} data, but the MIRI spectrum reveals many more lines, and the greatly increased resolving power allows the shapes of the mid-infrared lines and hence the dynamics of the atomic and warm molecular gas, to be analyzed in detail for the first time. 

Of particular note is that the NGC 7469 nuclear spectrum shows a striking lack of strong PAH emission, and an abundance of high ionization lines, tracing the hard ionizing radiation field emerging from the AGN.  There are also eight H$_{2}$ emission lines present in the nuclear spectrum indicative of warm molecular gas. Finally, there is a slight blue upturn in the nuclear spectrum shortward of about $8\mu$m that may signal the presence of very warm dust heated by the AGN.  The comparison of the two \emph{JWST/MIRI} spectra demonstrates that in the total spectrum, as in the \emph{Spitzer/IRS} spectrum, the PAH emission is dominated by the starburst ring, while most of the high-ionization lines (not surprisingly) are coming from the nucleus. The H$_{2}$ emission emerges from the ring and the nucleus.  The properties of the outflow and the dust and gas in the starburst ring as traced with MIRI, are discussed in \cite{u22} and \cite{lai22}, respectively.  Here we concentrate on the properties of the high resolution nuclear spectrum and what they reveal about the AGN and the wind.

\begin{figure}
	\centering
	\includegraphics[width=0.47\textwidth]{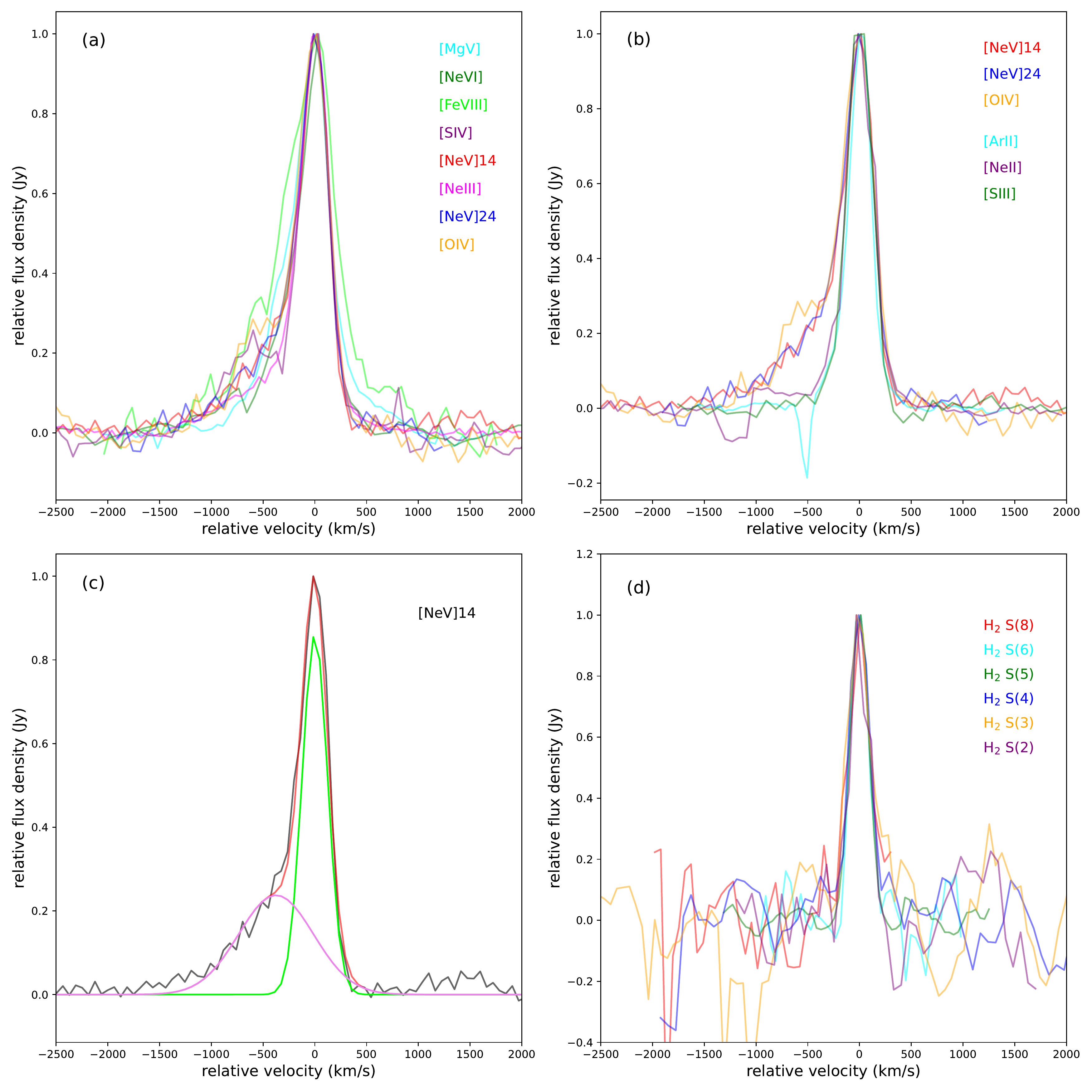}
	\caption{Atomic and H$_{2}$ line profiles in the MIRI/MRS nuclear spectrum of NGC 7469. High-ionization lines are shown in (a).  A comparison of some of the brighter high-ionization and low-ionization lines is shown in (b). All lines have been normalized and shifted to the centroid of the narrow line peak. The high ionization lines are all asymmetric, with wings that extend up to $1700$ km s$^{-1}$ to the blue. The high velocity, blue wings are not present in the low-ionization lines. A simple two-component Gaussian fit to the $14.3\mu$m [NeV] line is shown in (c). A number of the H$_{2}$ emission lines detected in the NGC 7469 nuclear spectrum are shown in (d) for comparison.  While the H$_{2}$ lines are resolved, they are all symmetric and narrow.} 
	\label{FIG:profiles}
\end{figure}

\subsection{Ionized Gas}

\begin{figure}[t!]
	\centering
	\includegraphics[width=0.47\textwidth]{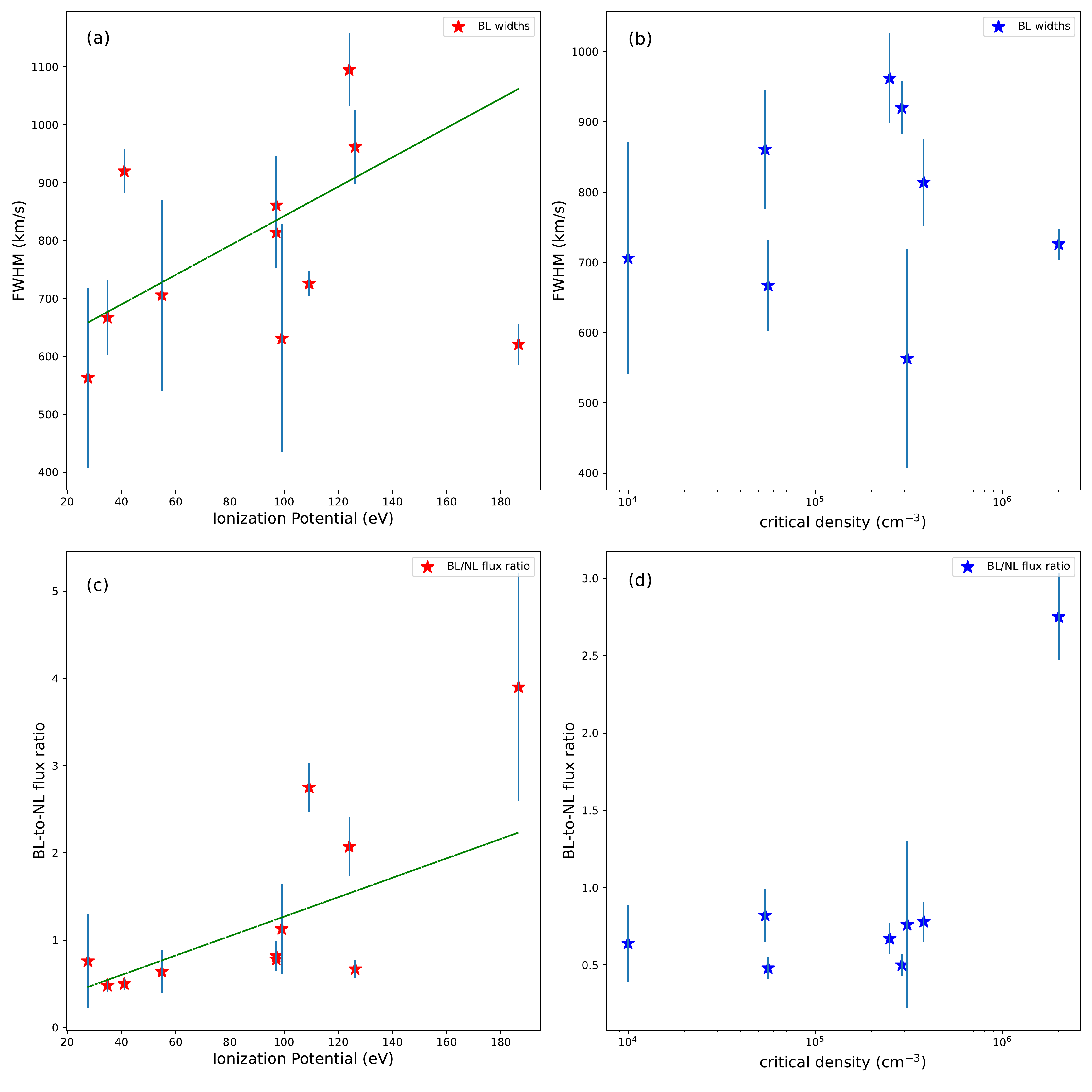}
	\caption{Emission line properties as a function of ionization and critical density in NGC 7469.  The broad line widths and broad-to-narrow component flux ratios are plotted as a function of ionization potential (a,c) and critical density (b,d) for the fine structure lines in the nuclear spectrum of NGC 7469.  The lines show a correlation of broad line width and broad to narrow line flux ratio with ionization potential. Linear fits are shown as green, dot-dashed lines in panels (a) and (c), excluding the [MgVII] $5.5\mu$m line which is blended with H$_{2}$.}
	\label{FIG:ionization}
\end{figure}

We detect emission lines from the nucleus of NGC 7469 spanning a wide range of ionization states, from [FeII] with IP=7.9 eV to [MgVII] with IP=186.5 eV, allowing us to explore properties of the circumnuclear atomic gas as a function of ionization. The coronal lines are strong in the nucleus of NGC 7469. The [NeV]/[NeII] and [OIV]/[NeII] line flux ratios are 1.5, and 5.0, respectively, comparable to the values seen in nearby AGN and AGN-dominated ULIRGs with $ISO$ and \emph{Spitzer} \citep{genzel98,Lutz00,Sturm02,armus04,weedman05,armus07}.

Because of the high spectral resolving power of MIRI, nearly all of the emission lines are resolved in the MRS nuclear spectrum.  All lines show a strong narrow component that dominates the emission, while some show a more complex line profile. The narrow line widths, corrected for instrumental broadening, range from $\sim100-400$ km s$^{-1}$, with a median width of about 260 km s$^{-1}$. While the wavelength calibration is still being refined for MRS, in most cases the centroids of the narrow line components are consistent with the systemic velocity of NGC 7469. 

A large number of the emission line profiles exhibit broad, blue wings.  This broad emission is prominent in the high-ionization, coronal lines (e.g. [NeV], [NeVI], [MgVII]) but absent from the low-ionization lines (e.g., [NeII]). The high velocity emission extends up to $1700$ km s$^{-1}$ blueward of the narrow component. There is little or no redshifted emission in the high-ionization lines, with the exception of [MgV] $5.61\mu$m and [FeVIII] $5.447\mu$m, which do show red wings extending out to about 1000 km s$^{-1}$. We have fit all lines that show broad emission with a simple two-component Gaussian model profile to extract and compare the basic parameters of the components between lines. The broad components have FWHM that range from $\sim500 - 1000$ km s$^{-1}$. While we designate these features as broad to facilitate a comparison to the narrow emission, they are far narrower than those from the canonical broad line region.  The FWHM of the H$\beta$ broad line in NGC 7469 is over 4300 km s$^{-1}$ \citep{peterson04}. The offsets between the centroids of the narrow and broad components range from $\sim100 - 600$ km s$^{-1}$. The fluxes and widths of the narrow and broad components are presented in Table 1 and a subset of the lines are shown scaled and shifted to a common velocity in Fig.~\ref{FIG:profiles}. Because the line is very bright, the broad blueshifted emission in the [OIV] line was also seen in the \emph{Spitzer/IRS} long-high spectrum of NGC 7469 \citep{Inami13}, even though the spectral resolving power was only $\sim550$ and the slit was extremely wide (including emission from the disk and starburst ring). The [NeIII] and [NeV] $14\mu$m lines were also resolved in the \emph{Spitzer} spectra, with intrinsic widths of $\sim420$ and $\sim350$ km s$^{-1}$, respectively.

\subsection{Warm Molecular Gas}

The MIRI MRS spectrum of the nucleus of NGC 7469 reveals the presence of a number of pure rotational emission lines of H$_{2}$ that trace the warm molecular gas in the vicinity of the AGN.  Eight transitions are detected in the spectrum, with fluxes ranging from $\sim0.5-5 \times 10^{-18}$ W m$^{-2}$.  All the H$_{2}$ lines are resolved, with intrinsic line widths (FWHM) ranging from $124-331$ km s$^{-1}$.  The median line width is $\sim210$ km s$^{-1}$ FWHM. None of the H$_{2}$ emission lines show the broad, blueshifted emission seen in the high ionization lines (see Fig. ~\ref{FIG:profiles}). The widths of the H$_{2}$ lines are comparable to, or less than, those of the low ionization atomic lines.  The S(1) line is the broadest, but its width is highly uncertain. 

Following \cite{togi06}, we have fit the H$_{2}$ line fluxes with a power-law, temperature dependent excitation function to estimate the characteristic gas temperature and implied mass. The power-law index is 4.55, and there is approximately $1.3\times10^{5}$ M$_{\odot}$ of warm gas at T $\geq$ 200K, and an extrapolated mass (down to 50K) of $\sim1.2\times10^{7}$ M$_{\odot}$ in a 0.3\arcsec diameter aperture.  The latter is approximately 0.4\% of the total estimated H$_{2}$ mass within a radius of 2.5\arcsec in NGC 7469 \citep{Davies04, Izumi20} and 18\% the estimated H$_{2}$ mass within a radius of 0.2\arcsec \citep{Izumi20}.

\subsection{Dust}

The strong PAH features that dominate the total mid-infrared spectrum of NGC 7469 are weak or absent from the nuclear spectrum (see Fig. \ref{FIG:FullSpectra}).  The spectrum does have some residual continuum noise that makes identification of very weak and broad PAH features difficult, but it is clear that the strong PAH emission seen in the total MIRI and \emph{Spitzer} spectra is not present in the nuclear spectrum.  This is consistent with the results of \cite{lai22} and  \cite{honig2010}. We do, however, detect a weak PAH feature at $6.2\mu$m with a flux of $\sim2.7\pm0.2\times10^{-17}$ W m$^{-2}$, and $2.5\sigma$ upper limits on the fluxes of the $7.7\mu$m and $11.3\mu$m PAH features of $\leq 9.2\times10^{-17}$ W m$^{-2}$, and $\leq 3.5\times10^{-17}$ W m$^{-2}$, respectively.  The $6.2\mu$m nuclear flux is about 0.5\% of the PAH emission in the ``total" MIRI spectrum. Our 6.2/7.7 PAH flux ratio limit of $\ge0.11$ does not allow us to confidently constrain the grain size, but the $6.2\mu$m PAH Equivalent Width (EQW) of $0.07\mu$m is within the range seen for other AGN \citep[e.g.,][]{spoon07}. 

The NGC 7469 nuclear spectrum is steeply rising over much of the MIRI wavelength range with a slight inflection around $\sim8\mu$m that may signal the presence of warm dust heated by the AGN. The 15 to $5\mu$m flux density ratio is $\sim12.5$.  There is no indication of either strong silicate absorption or emission at $9.7\mu$m or $18\mu$m.  The $9.7\mu$m silicate strength, as defined in \cite{hao2007}, is $0.02$. This is significantly smaller than found for many nearby QSOs where silicate emission is common, larger than most Seyfert 2's and ULIRGs where absorption dominates and the silicate strengths are negative, but consistent with other Seyfert 1 galaxies observed with \emph{Spitzer} \citep{hao2005, hao2007, schweitzer06, netzer07}. For comparison, the silicate strength measured in the \emph{Spitzer} spectrum of NGC 7469 is -0.14, indicative of a small apparent absorption, undoubtedly the result of a significant amount of optically thick dust from the starburst ring within the large \emph{Spitzer} slit.  The silicate strength together with the $6.2\mu$m PAH EQW places NGC 7469 between Class 1A and 1B objects in the diagnostic diagram of \cite{spoon07}. The low silicate strength, weak PAH EQW and steep spectral slope may also indicate a contribution from ongoing star formation to the nuclear spectrum \citep{marshall18}.

\section{Discussion}

The high spectral resolution of the MIRI MRS data, together with the range of detected atomic and H$_{2}$ lines allows us to probe the dynamics and ionization structure of the atomic gas and the warm molecular gas in NGC 7469 on scales of $\sim100$pc in the mid-infrared for the first time. 

NGC7469 is known to have an outflowing, highly-ionized nuclear wind seen in the near-infrared [SiVI] coronal line on sub-arcsecond scales \citep{Muller-Sanchez11}. The atomic gas measured with MIRI is most likely associated with this outflow. 
One striking difference, however, is that the MIRI data show gas moving at much higher velocity than mapped in [SiVI], which reaches velocities of $\pm120$km s$^{-1}$ at radii of about $0.2-0.3$\arcsec. Broad blueshifted emission extending to nearly $\sim1000$ km s$^{-1}$ is seen in the central spaxel of the Keck/OSIRIS data, but the single Gaussian fits used to construct the outflow model are clearly dominated by slower moving gas at radii of $\sim60$ pc. The high-ionization, high velocity gas in the \emph{JWST} MIRI spectrum is also seen as a faint, Eastern extension in the [MgV] map produced by \cite{u22}. 

The fact that the high velocity gas is most prominent in the coronal lines is consistent with an ionized, decelerating wind. In this picture, the more highly ionized gas is closer to the AGN and is moving at greater speeds. The enhanced blueshifted emission is a result of geometry and possibly extinction, with the receding gas being partially blocked by dust in the inner disk. Despite the high velocities, the bulk of the gas appears to be photoionized, as the [SIV]/[NeII], [NeIII]/[NeII] and the limit on the [FeII]/[OIV] ratios all imply that shock heating is negligible \citep{Inami13}. It is likely that the coronal line gas we see in NGC 7469 may also be related to the high velocity, warm absorber and emission line gas inferred to exist on much smaller scales ($0.1-3$ pc) by \cite{Grafton-Waters20} through modeling of the XMM/RGS spectrum. Blueshifted coronal lines in the mid-infrared have been seen in nearby ULIRGs with powerful outflows \citep[e.g.,][]{spoon09}, and it is reminiscent of the kinematic structures seen in some Narrow-Line Seyfert Type-1 galaxies (NLSI's) known as ``blue outliers" \citep{komossa08}, typically identified via asymmetric optical [OIII] emission. 

To better understand the NGC 7469 outflow probed with MIRI, we plot the broad line widths and the broad-to-narrow line flux ratios as a function of ionization potential and critical density in Fig.~\ref{FIG:ionization}. With the exception of the [MgVII] line at $5.503\mu$m which is partially blended with H$_{2}$ S(7), the broad line widths show a moderate correlation with ionization potential and the broad to narrow line flux ratio (Pearson's r=0.58 and r=0.57, respectively). There is no correlation with critical density, but this is not surprising since the ratio of the [NeV] $14\mu$m and [NeV] $24\mu$m lines is $\sim0.96$ ($\sim0.94$ for the broad lines alone), indicating the coronal line gas is in the low density limit with n$_{e} \leq 300$cm$^{-3}$ \citep[see][]{alexander99}. The primary correlations of the line widths and relative fluxes are with ionization, similar to what is seen in NGC 1068 \citep{Lutz00}. The correlations among fit components strengthen the picture of the simple model that is evident from the line profiles alone, namely that the prominence of the high velocity, mostly blueshifted gas among the coronal lines seen with MIRI suggests a decelerating, stratified, AGN driven outflow in NGC 7469.  

To estimate the mass outflow rate in the coronal gas, we need to know the gas density, filling factor, maximum velocity and lateral surface area of the wind.  From the [SiVI] emission, \cite{Muller-Sanchez11} estimate a mass outflow rate of $\sim4$ M$_{\odot}$ yr$^{-1}$, assuming a density of 5000 cm$^{-3}$, and a filling factor of 0.001 together with a modeled wind area of $11\times10^{4}$ pc$^{2}$ and a maximum velocity of 130 km s$^{-1}$. If the high velocity gas we see with \emph{JWST} emerges in the wind over an area consistent with our aperture and the Eastern extension in \cite{u22}, and the gas has a maximum velocity of 1700 km s$^{-1}$, a density of $\sim300$ cm$^{-3}$, and a filling factor of 0.001, then we would arrive at a mass outflow rate of $\sim0.5-1$ M yr$^{-1}$, significantly smaller than that estimated by \cite{Muller-Sanchez11}. The mass outflow rate is highly uncertain, but is still much larger that the gas accretion rate required to power the AGN.  Using the scaling between [OIV] and the Black Hole Accretion Rate (BHAR) in \cite{stone22}, we estimate a BHAR of $\sim4.1\times10^{-2}$ M$_{\odot}$yr$^{-1}$, or roughly one to two orders of magnitude less than the mass outflow rate in the ionized wind.  

As has been shown by \cite{komossa07}, the narrow line widths can be used as a surrogate for the stellar velocity dispersion and therefore used to estimate the black hole mass in Seyfert galaxies, once components associated with any outflowing gas are removed. While many of the lines show clear blueshifted emission in the MIRI MRS spectrum of NGC 7469, the narrow components have centroids consistent with the systemic velocity.  This strong narrow line component can be used to estimate the mass of the central SMBH, assuming it dominates the dynamical mass in the sampled volume.  

Correlations between the widths and luminosities of the mid-infrared [NeV] $14.3\mu$m and [OIV] $25.9\mu$m lines and the mass of the central SMBH have been established by \cite{dasyra08} using \emph{Spitzer/IRS} data of a sample nearby Seyfert galaxies, including NGC 7469. Using the width of the narrow component of the [NeV] and [OIV] lines in the MIRI/MRS nuclear spectrum, we estimate a mass of $\sim6\times10^{6}$ M$_{\odot}$, about half the value calculated from reverberation line estimates \citep{peterson04}, or one third the value estimated by \cite{Nguyen21} from fits to the rotating central nuclear disk measured in [CI] and CO with \emph{ALMA}.  Similarly, the luminosities of the narrow lines imply SMBH masses of 4.4 and $4.8\times10^{7}$ M$_{\odot}$ for [NeV] and [OIV], respectively, or about a factor of 4-5 larger than the reverberation line esimate. As noted by \cite{dasyra08}, a large Eddington ratio would lower the SMBH mass predicted from the [OIV] and [NeV] line luminosities perhaps bringing them more in line with the reverberation line estimate. The widths of the broad components in the MIRI/MRS nuclear spectrum imply SMBH masses that are a factor of 75-80 higher than measured by \cite{peterson04}. This is consistent with our association of the high velocity gas with the outflow. Future studies of large samples of Seyfert galaxies with \emph{JWST}, where outflowing and infalling gas can be identified and separated with the MIRI and NIRSpec IFUs, could provide a revised calibration of this correlation and a powerful tool to estimate SMBH masses in Seyfert galaxies.

The strong but relatively narrow H$_{2}$ lines in the nuclear spectrum along with the relatively flat power law index suggest that warm molecular gas in the central $\sim100$ pc is dominated by AGN heating.  This is consistent with the relative dip in the line dispersion seen by U et al. (2022) on the nucleus. The fan-shaped structure to the Northwest seen as an enhancement in the H$_{2}$ dispersion by \cite{u22} may reveal molecular gas shock heated by the wind. The enhanced central [CI] emission has also been suggested as evidence for AGN heating of the circumnuclear gas in an XDR \citep{Izumi20}. The near-infrared H$_{2}$ line ratios can be explained by AGN heating or arise in dense gas illuminated by hot stars \citep{davies05}. If the $6.2\mu$m PAH emission seen in the nuclear spectrum is powered by star formation, then the implied star formation rate is $\sim0.4$ M$_{\odot}$ yr$^{-1}$ \citep{pope08}, or less than 1\% of the total star formation rate in NGC 7469.


\section{Summary}

We provide an analysis of the high-resolution, mid-infrared spectrum of the nucleus of NGC 7469 taken with the MIRI instrument onboard the \emph{James Webb Space Telescope}.  The rich set of emission features have provided a detailed picture of the dynamics and physical conditions of the ionized atomic and warm molecular gas, and the dust in inner $\sim100$ pc of NGC 7469. Our results can be briefly summarized as follows:

\begin{enumerate}

\item 
There are 19 identified emission lines covering a wide range of ionization potential up to 187 eV.  The high ionization lines all show broad, blueshifted emission reaching up to -1700 km s$^{-1}$ with respect to bright, narrow systemic emission. The broad, blueshifted emission is not seen in the low-ionization lines.  The width of the broad emission and the broad to narrow line flux ratios correlate with ionization potential.  We interpret the results as indicative of a decelerating, stratified, AGN driven outflow emerging from the NGC 7469 nucleus. The estimated mass outflow rate in the wind is significantly larger than the current black hole accretion rate needed to power the AGN.

\item
There are eight pure rotational H$_{2}$ emission lines detected in the nuclear spectrum. The H$_{2}$ lines have a median intrinsic FWHM $\sim210$ km s$^{-1}$, with none showing the broad, blueshifted emission seen in the high ionization atomic lines. We estimate there is a total mass of warm gas of $\sim1.2\times 10^{7}$ M$_{\odot}$ in the central 100pc. The high temperature and relatively narrow H$_{2}$ lines in the nuclear spectrum suggest that this warm molecular gas is heated by the AGN.

\item 
The PAH features that dominate the total mid-infrared spectrum of NGC 7469 are significantly weaker in the nuclear spectrum.  We detect a $6.2\mu$m feature with an EQW $\sim0.07\mu$m and a flux of $2.7\times 10^{-17}$ Wm$^{-2}$.  If the PAH emission in the nuclear spectrum is a result of star formation, then the implied star formation rate is $\sim0.4$ M$_{\odot}$ yr$^{-1}$, or less than 1\% of the total star formation rate in NGC 7469.

\item
The NGC 7469 nuclear spectrum is steeply rising over the MIRI wavelength range. There is no indication of strong silicate absorption or emission at $9.7\mu$m. The measured silicate strength is $0.02$, significantly smaller than found for many nearby QSOs, but consistent with other Seyfert 1's.

\end{enumerate}

The mid-infrared, nuclear spectrum of NGC 7469 presented here demonstrates the power of MIRI/MRS to disentangle the complex excitation and kinematic properties in the circumnuclear environments of nearby merging galaxies that are both feeding AGN and rapidly producing stars. These regions are inherently multi-phase, and the ability to spatially and spectrally resolve the atomic and molecular gas emission as well as the dust in these galaxies is a unique ability made possible with infrared integral field spectrographs on \emph{JWST}. Planned and future \emph{JWST} observations of NGC 7469 and other nearby LIRGs and Seyfert galaxies will undoubtedly shed even more light on feeding and feedback at the centers of rapidly evolving galaxies.


\section*{Acknowledgements}

This work is based on observations made with the NASA/ESA/CSA \emph{JWST}. The research was supported by NASA grant JWST-ERS-01328. The data were obtained from the Mikulski Archive for Space Telescopes at the Space Telescope Science Institute, which is operated by the Association of Universities for Research in Astronomy, Inc., under NASA contract NAS 5-03127 for JWST. VU acknowledges funding support from NASA Astrophysics Data Analysis Program (ADAP) grant 80NSSC20K0450. The Flatiron Institute is supported by the Simons Foundation. AMM acknowledges support from the National Science Foundation under Grant No. 2009416. ASE and SL acknowledge support from NASA grant HST-GO15472. YS was funded in part by the NSF through the Grote Reber Fellowship Program administered by Associated Universities, Inc./National Radio Astronomy Observatory. F.M-S. acknowledges support from NASA through ADAP award 80NSSC19K1096. SA gratefully acknowledges support from an ERC Advanced Grant 789410, from the Swedish Research Council and from the Knut and Alice Wallenberg (KAW) Foundation. KI acknowledges support by the Spanish MCIN under grant PID2019-105510GB-C33/AEI/10.13039/501100011033.
Finally, this research has made use of the NASA/IPAC Extragalactic Database (NED) which is operated by the Jet Propulsion Laboratory, California Institute of Technology, under contract with the National Aeronautics and Space Administration.

\vspace{5mm}

\facilities{\emph{JWST} (NIRCam and MIRI)}


\software{astropy~\citep{2013A&A...558A..33A,2018AJ....156..123A},
Cosmology calculator~\citep{Wright06},
\emph{JWST} Science Calibration Pipeline~\citep{jwstpipe},
CAFE~\cite[][D\`iaz-Santos et al. 2022, in preparation]{Marshall07}, 
JDAVis~\citep{Lim22},
lmfit \citep{Newville2014},
Matplotlib \citep{Hunter2007},
Numpy \citep{VanderWalt2011},
QFitsView \citep{Ott2012},
SciPy \citep{Virtanen2020}}



\bibliography{n7469nuc}{}
\bibliographystyle{aasjournal}



\end{document}